# Delayed treatment with nimesulide reduces measures of oxidative stress following global ischemic brain injury in gerbils

Authors: Eduardo Candelario-Jalil *; Dalia Alvarez; Nelson Merino and Olga Sonia León

Affiliation: Department of Pharmacology, University of Havana (CIEB-IFAL), Apartado Postal 6079, Havana City 10600, Cuba.

*Author to whom all correspondence should be addressed:

**Eduardo Candelario-Jalil, Ph.D.**

**Department of Pharmacology**

**University of Havana (CIEB-IFAL)**

**Apartado Postal 6079**

**Havana City 10600**

**CUBA**

**Tel.: +53-7-271-9534**

**Fax: +53-7-336-811**

**E-mail: candelariojalil@yahoo.com**

**Abstract**

Metabolism of arachidonic acid by cyclooxygenase is one of the primary sources of reactive oxygen species in the ischemic brain. Neuronal overexpression of cyclooxygenase-2 has recently been shown to contribute to neurodegeneration following ischemic injury. In the present study, we examined the possibility that the neuroprotective effects of the cyclooxygenase-2 inhibitor nimesulide would depend upon reduction of oxidative stress following cerebral ischemia. Gerbils were subjected to 5 min of transient global cerebral ischemia followed by 48 h of reperfusion and markers of oxidative stress were measured in hippocampus of gerbils receiving vehicle or nimesulide treatment at three different clinically relevant doses (3, 6 or 12 mg/kg). Compared to vehicle, nimesulide significantly ($P<0.05$) reduced hippocampal glutathione depletion and lipid peroxidation, as assessed by the levels of malondialdehyde (MDA), 4-hydroxy-alkenals (4-HDA) and lipid hydroperoxides levels, even when the treatment was delayed until 6 h after ischemia. Biochemical evidences of nimesulide neuroprotection were supported by histofluorescence findings using the novel marker of neuronal degeneration Fluoro-Jade B. Few Fluoro-Jade B positive cells were seen in CA1 region of hippocampus in ischemic animals treated with nimesulide compared to vehicle. These results suggest that nimesulide may protect neurons by attenuating oxidative stress and reperfusion injury following the ischemic insult with a wide therapeutic window of protection.

**Acknowledgements:**

The authors are greatly indebted to Dr. Larry C. Schmued (Division of Neurotoxicology, National Center for Toxicological Research, Food and Drug Administration, Jefferson, Arizona, USA) for providing Fluoro-Jade B for these studies. The expert technical assistance of Noël H. Mhadu is greatly acknowledged. We are also grateful to Dr. Gregorio Martínez-Sánchez for fruitful discussions.

## 1. Introduction

Neuronal loss following ischemic brain injury is a delayed process in which the primary injury is followed by a prolonged period of secondary neurodegeneration resulting in neurological dysfunction. It has been shown that pyramidal neurons in the hippocampal CA1 region are particularly sensitive to injury due to transient global ischemia, and neuronal cell death occurs within days after ischemia/reperfusion (Kirino, 1982; Schmidt-Kastner and Freund, 1991). This phenomenon is referred to as delayed neuronal death (DND). Although not fully understood, ischemia-induced DND is probably associated with a myriad of biochemical events initially triggered by the extracellular accumulation of glutamate. In turn, this leads to membrane depolarization, increased concentrations of intracellular calcium, overproduction of reactive oxygen species and oxidative damage (Kitagawa et al., 1990; Nakamura et al., 1999; Urabe et al., 2000; Chan, 2001).

Many factors, including high content of polyunsaturated fatty acids, high rate of oxidative metabolic activity, intense production of reactive oxygen metabolites, relatively low antioxidant capacity and low repair mechanism activity, suggest that neurons in the central nervous system are particularly prone to oxidative stress (Halliwell and Gutteridge, 1985; Liu, 2003). A large body of experimental evidences has demonstrated the involvement of reactive oxygen species in cerebral ischemia/reperfusion and cell death mechanisms (Kitagawa et al., 1990; Facchinetti et al., 1998; Chan, 2001).

Previous studies have shown that one of the primary sources of reactive oxygen species in the ischemic brain is through the metabolism of arachidonic acid by cyclooxygenase (Nelson et al., 1992; Busija et al., 1998; Chan, 2001). Two isoforms of cyclooxygenase have been described: cyclooxygenase-1 and cyclooxygenase-2. Cyclooxygenase-1 is expressed constitutively in many organs and contributes to the synthesis of prostanoids involved in normal cellular functions (Seibert et al., 1997). Cyclooxygenase-2 is thought to be an inducible enzyme, the expression of which is up-regulated in inflammation (Seibert et al., 1997). Recently, cyclooxygenase-2 mRNA and protein levels have been shown to be significantly increased within neurons and vascular cells after cerebral ischemia and other insults that result in neurodegeneration (Nogawa et al., 1997; Nakayama et al., 1998; Koistinaho et al., 1999; Iadecola et al., 1999). Cyclooxygenase-2 has become the focus of attention because it is the rate-limiting enzyme involved in arachidonic acid metabolism, thereby generating prostaglandins and thromboxanes, molecules that play important roles in supporting and sustaining the inflammatory response seen after cerebral ischemia (Vane et al., 1998; Iadecola and Alexander, 2001).

It is now well known that the administration of cyclooxygenase-2 selective inhibitors is able to confer protection against brain injury in relevant models of cerebral ischemia even when the treatment is delayed until several hours after ischemia (Nogawa et al., 1997; Candelario-Jalil et al., 2002a,b), suggesting that cyclooxygenase-2 activity is involved in the biochemical events that lead to delayed neuronal cell death.

Nimesulide (N-(4-nitro-2-phenoxyphenyl)-methanesulfonamide) is a non-steroidal anti-inflammatory drug with potent effects, showing a high affinity and selectivity for cyclooxygenase-2 (Cullen et al., 1998). Nimesulide readily crosses the intact blood-brain barrier in both humans and rodents (Taniguchi et al., 1997; Cullen et al., 1998). Several recent studies have demonstrated a marked neuroprotective effect of nimesulide on chronic cerebral hypoperfusion (Wakita et al., 1999), kainate-induced excitotoxicity (Candelario-Jalil et al., 2000), quisqualic acid-induced neurodegeneration (Scali et al., 2000), diffuse traumatic brain injury (Cernak et al., 2001, 2002), glutamate-mediated apoptotic damage (Mirjany et al., 2002) and induction of the expression of the B subunit of endogenous complement component C1q (C1qB) in transgenic mice with neuronal overexpression of human cyclooxygenase-2 (Spielman et al., 2002).

Recently, we have found a significant neuroprotective effect of nimesulide in both global (Candelario-Jalil et al., 2002b) and focal cerebral ischemia (Candelario-Jalil et al., 2002c). In addition, previous results from our lab indicated that nimesulide is able to reduce hippocampal oxidative damage following excitotoxic brain injury induced by the systemic application of kainate (Candelario-Jalil et al., 2000). However, the effects of nimesulide on measures of oxidative stress in a model of cerebral ischemia had not been previously investigated.

In the light of all these evidences and given that the activation of the cyclooxygenase pathway by cerebral ischemia, especially during reperfusion, accelerates conversion of arachidonic acid to prostaglandins and thromboxanes, producing reactive oxygen species such as hydroxyl radicals and superoxide radicals (Katsuki and Okuda, 1995; Miettinen et al., 1997), the purpose of this study was to examine the effects of clinically-relevant doses of nimesulide on measures of oxidative injury seen in the hippocampus after transient forebrain ischemia in gerbils.

## 2. Materials and Methods

### *2.1. Transient forebrain ischemia*

Studies were performed in accordance with the Guide for the Care and Use of Laboratory Animals as adopted and promulgated by National Institutes of Health (Bethesda, MD, USA). Our institutional animal care and use committee approved the experimental protocol (No. 02/112). Adult male Mongolian gerbils (*Meriones unguiculatus*, 12-15 weeks, 60-70 g) were subjected to transient global cerebral ischemia under diethyl ether anesthesia by occluding both common carotid arteries for 5 min with microaneurysmal clips (Sugita, Japan) exactly as in our previous reports (Candelario-Jalil et al., 2001, 2002a,b; Martínez et al., 2001), which consistently resulted in delayed neuronal death in the CA1 region of the hippocampus (Kirino, 1982; Candelario-Jalil et al., 2002a,b). Blood flow during the occlusion and reperfusion after removal of the clips was visually confirmed and the incision was closed with 4-0 silk sutures. The onset of cerebral ischemia was associated with a brief period of panting breathing and body movements followed by quiescence. Successful occlusion of both carotid arteries was evident with the rapid onset of complete bilateral ptosis and the adoption of a ‘hunched’ posture. In sham-operated animals, the arteries were freed from connective tissue but were not occluded. The rectal temperature was carefully monitored and maintained at 37 ± 0.5°C using an incandescent lamp and the animals were allowed to recover on an electrical heated blanket. In addition, rectal temperature was monitored at 6-h intervals until sacrificing the animals in all experimental groups.

### *2.2. Drug treatments*

The following experimental groups were prepared: a sham-operated group (n=7), an ischemic group treated with the vehicle of nimesulide (polyvinylpyrrolidone, n=9) and three groups of ischemic gerbils treated with nimesulide at doses of 3 mg/kg (n=7), 6 mg/kg (n=9) and 12 mg/kg (n=9) given intraperitoneally starting after 6 h of reperfusion (delayed treatment schedule). Additional doses of nimesulide or vehicle were administered at 12, 24 and 36 h of reperfusion. These doses and treatment paradigm were based on our previous results showing marked neuroprotective efficacy of nimesulide in this gerbil model of cerebral ischemia (Candelario-Jalil et al., 2002b). Three different groups of sham-operated animals (n=7 each) treated with nimesulide using the same doses (3, 6 and 12 mg/kg) and treatment schedule of those ischemic animals treated with the inhibitor were also included. We also performed an experiment in which nimesulide (6 mg/kg) was administered 30 minutes before ischemia and again at 6, 12, 24 and 36 h of reperfusion (pre-treatment schedule, n=9). A vehicle pre-treated group was also included (n=7).

### *2.3. Brain sampling*

Animals were sacrificed after 48 h of reperfusion, because according to our previous results, hippocampal oxidative damage (significant depletion in reduced glutathione and a marked increase in oxidized glutathione, malondialdehyde and 4-hydroxy-alkenals levels) is maximal at this time point in gerbil global cerebral ischemia (Candelario-Jalil et al., 2001). Gerbils were deeply anesthetized with diethyl ether and perfused transcardially with ice-cold saline to flush all blood components from the vasculature. Brains were quickly removed, kept in ice-cold saline and hippocampi were immediately dissected out on a chilled plate, exactly as in our previous reports (Candelario-Jalil et al., 2000, 2001; Martínez et al., 2001). Hippocampi were weighed and homogenized in ice-cold 20 mM Tris-HCl buffer (pH 7.4) and centrifuged for 10 min at 12 000 *g*. The supernatant was collected, frozen at -70°C and employed for biochemical analyses.

### *2.4. Glutathione determination*

Reduced and oxidized glutathione (GSH and GSSG, respectively) were measured enzymatically in 5-sulphosalycilic acid-deproteinized samples by using a modification (Anderson, 1985) of the procedure of Tietze (1969) as described for brain samples (Floreani et al., 1997). The method is based on the determination of a chromophoric product, 2-nitro-5-thiobenzoic acid, resulting from the reaction of 5,5'-dithiobis(2-nitrobenzoic acid) with GSH. In this reaction, GSH is oxidized to GSSG, which is then reconverted to GSH in the presence of glutathione reductase (type III, from *Saccharomyces cerevisiae*, Randox Laboratories, Antrim, UK) and NADPH. The rate of 2-nitro-5-thiobenzoic acid formation, which is proportional to the sum of GSH and GSSG present, is followed at 412 nm. Samples were assayed rapidly to minimize GSH oxidation. Specificity of this method for glutathione quantification is ensured by highly specific glutathione reductase. GSH present in the samples was calculated as the difference between total glutathione and GSSG levels, taking into account the fact that one molecule of GSSG gives rise to two molecules of GSH upon reaction with glutathione reductase. A standard curve with known amounts of GSH was established and employed for estimating glutathione content.

### *2.5. Lipid peroxidation assays*

Lipid peroxidation was assessed by measuring the concentration of malondialdehyde (MDA) and 4-hydroxy-alkenals (4-HDA) and by determining the levels of lipid hydroperoxides in brain samples. Concentrations of MDA and 4-HDA were analyzed using the LPO-586 kit obtained from Calbiochem (La Jolla, CA, USA). In the assay, the production of a stable chromophore after 40 min of incubation at 45°C was measured at a wavelength of 586 nm. For standards, freshly prepared solutions of malondialdehyde bis [dimethyl acetal] (Sigma) and 4-hydroxy-nonenal diethylacetal (Cayman Chemical, Ann Arbor, MI, USA) were employed and assayed under identical conditions. Concentrations of MDA and 4-HDA in brain samples were calculated using the corresponding standard curve and values were expressed as nmol MDA+4-HDA per mg protein. This procedure has been used widely for the measurement of products of lipid peroxidation in brain homogenates (Chabrier et al., 1999; Candelario-Jalil et al., 2000, 2001). Lipid hydroperoxides were measured with ferrous oxidation-xylenol orange assay (Gay et al., 1999) as reported for brain homogenates (Song et al., 1999; Candelario-Jalil et al., 2001). Hydrogen peroxide was used as reference standard (R=0.997). Lipid hydroperoxides levels were expressed as nmol hydroperoxides per mg protein.

*2.6. Fluoro-Jade B staining*

We also performed histochemical analyses of brain samples obtained from ischemic animals treated with nimesulide in order to confirm the neuroprotective effect of this compound against hippocampal CA1 neuronal loss following transient global ischemia (Candelario-Jalil et al., 2002b). We employed the novel high affinity fluorescent marker Fluoro-Jade B for the localization of neuronal degeneration following ischemia (Schmued and Hopkins, 2000). Three different groups of gerbils were prepared: a sham-operated group (n=5), a vehicle-treated ischemic group (n=7) and a nimesulide-treated group (6 mg/kg; i.p., starting after 6 h of reperfusion, n=7). Taking into account that previous studies have found marked neuroprotective effects with nimesulide at the dose of 6 mg/kg in different models of brain injury (Candelario-Jalil et al., 2000, 2002b; Cernak et al., 2001, 2002), we decided to select this dose for this experiment. At 7 days after ischemia, animals were anesthetized with diethyl ether and then perfused transcardially with 0.1 M neutral phosphate buffered 10 % formalin (4 % formaldehyde). The brains were post-fixed overnight in the same fixative solution plus 20% sucrose. Brain tissue was cut on a freezing sliding microtome at a thickness of 25 μm. Three sections per animal containing the dorsal hippocampus were obtained at three levels (-1.0, -1.5 and -2.2 mm relative to bregma) and stained with Fluoro-Jade B for detection of degenerating neurons. Sections were mounted on 2% gelatin-coated slides and then air dried on a slide warmer at 50ºC for half an hour. The slides were first immersed in a solution containing 1% sodium hydroxide in 80% alcohol for 5 min. This was followed by 2 min in 70% alcohol and 2 min in distilled water. The slides were then transferred to a solution of 0.06% potassium permanganate for 10 min on a shaker table. The slides were then rinsed in distilled water for 2 min and transferred to a 0.0004% Fluoro-Jade B (Histochem, Jefferson, AR, USA) staining solution prepared in 0.1% acetic acid. After 20 min in the staining solution, the slides were rinsed for 1 min in each of three distilled water washes. Excess water was removed by briefly draining the slides vertically on a paper towel. The slides were then placed on a slide warmer, set at approximately 50ºC, until they were fully dry (aprox. 8 min). The dry slides were cleared by immersion in xylene for at least a min before coverslipping with DPX (Fluka, Milwaukee, WI, USA), a non-aqueous, non-fluorescent plastic mounting media. The slides were then examined using an epifluorescent microscope (Olympus, Tokyo, Japan) with blue (450-490 nm) excitation light and a barrier filter that allows passage of all wavelengths longer than 515 nm, resulting in a yellow emission color. With this method neurons that undergo degeneration brightly fluoresce in comparison to the background (Fig. 2) (Schmued and Hopkins, 2000). Images of the hippocampal CA1 *stratum pyramidale* area for both hemispheres were digitized using a CCD camera (Mavica, Sony, Japan). In each image, a rectangular box (0.25 x 1 mm) was centered over the CA1 cell layer beginning 1.0 mm lateral to the midline to maintain consistency across animals. An investigator who was uninformed of animal group identity, independently counted the number of fluorescing (positive) neurons in each brain image. The mean number of Fluoro-Jade B-positive neurons was calculated for each treatment group (Lyeth et al., 2001).

*2.7. Protein assay*

Total protein concentrations were analyzed using a Bio-Rad protein assay kit (Bio-Rad, Hercules, CA, USA). Analytical grade bovine serum albumin was used to establish a standard curve.

*2.8. Statistical analysis*

Data are expressed as mean ± standard deviation (S.D.). Statistical analysis was performed with one-way ANOVA followed by a Student-Newman-Keuls post-hoc test. The value of $P$ less than 0.05 was considered to be statistically significant.

## 3. Results

*3.1. Effects of nimesulide on ischemia-induced hippocampal glutathione depletion*

We have previously reported a significant late-onset (48-72 h) persistent glutathione depletion in hippocampus after global cerebral ischemia in gerbils (Candelario-Jalil et al., 2001). In the present study, the ANOVA revealed a significant effect of post-ischemic nimesulide treatment on glutathione homeostasis ($P<0.05$) compared to vehicle control (Table 1). Post-hoc analysis using the Student-Newman-Keuls test demonstrated that the three doses of nimesulide conferred a similar protection against the significant reduction in levels of GSH and increase in GSSG as shown in Table 1. Although there were no statistically significant differences among the nimesulide doses, in general, there was a trend toward a better neuroprotection with the highest doses (6 and 12 mg/kg). When nimesulide was administered before ischemia, no additional positive effects were noticed, as presented in Table 1. Administration of nimesulide to sham-operated animals failed to significantly modify hippocampal glutathione levels (Table 2).

*3.2. Effects of clinically relevant doses of nimesulide on measures of lipid peroxidation in ischemic hippocampus*

Malondialdehyde (MDA), 4-hydroxy-alkenals (4-HDA) and lipid hydroperoxides were used as indicators of lipid peroxidation in homogenates of hippocampus. In these studies, the ANOVA test resulted in a significant effect ($P<0.05$) of drug treatment on lipid peroxidation at 48 h following transient global cerebral ischemia. The Student-Newman-Keuls post-hoc analysis revealed that lipid peroxidation was significantly reduced by the delayed treatment with nimesulide compared to vehicle-treated ischemic animals (Fig. 1A and 1B). Similarly, no further protection was afforded by an additional drug treatment 30 min prior to ischemia (Fig. 1). On the other hand, administration of nimesulide (3, 6 or 12 mg/kg) to sham-operated animals failed to significantly modify lipid peroxidation markers (Table 2).

*3.3. Histofluorescence evidence of nimesulide neuroprotection against ischemia-induced hippocampal CA1 neuronal loss: Fluoro-Jade B staining*

The Fluoro-Jade B histofluorescence technique stained only degenerating neurons indicated by a bright fluorescence and was remarkably helpful in identifying neurodegeneration (Schmued and Hopkins, 2000). A representative Fluoro-Jade B labeling in the hippocampus is presented in Fig. 2. No Fluoro-Jade B positive fluorescence staining was noted in the hippocampal region of control animals (sham-operated control group) as shown in Fig. 2A. On the contrary, hippocampal sections from vehicle-treated ischemic gerbils showed a dramatic increase in Fluoro-Jade B positive cells in the CA1 pyramidal cell layer (Fig. 2B), whereas very few positive cells were seen in similar sections of the nimesulide-treated animals (Fig. 2C). Fig. 2D shows the quantitative analysis of Fluoro-Jade B-labeled neurons in the CA1 region of hippocampus in each treatment group. The neuroprotective effect of nimesulide as seen by Fluoro-Jade B (a specific marker of neuronal degeneration) supports our previous findings using a standard hematoxylin-eosin staining technique (Candelario-Jalil et al., 2002b).

The protection conferred by nimesulide in the present study is not attributable to effects on body temperature because this variable was carefully monitored and did not differ between treated and non-treated groups (data not shown).

## 4. Discussion

The results of the present study provide evidence that the neuroprotective effects of nimesulide are mediated, at least in part, through reduction of oxidative stress following transient global cerebral ischemia. In addition, these findings demonstrate that cyclooxygenase-2 selective inhibitors can reduce these oxidative events in the ischemic hippocampus, suggesting that cyclooxygenase-2 is involved in the late increase in measures of oxidative injury in the damaged brain.

Glutamate-induced excitotoxic damage, reactive oxygen species formation, and the generation of lipid peroxidation products (MDA, 4-HDA, lipid hydroperoxides) are prominent events thought to contribute to neuronal dysfunction and cell loss following traumatic damage and ischemic injury to the CNS (Facchinetti et al., 1998; Urabe et al., 2000; Chan, 2001). Sustained activation of glutamate receptors and the influx of $Ca^{2+}$ have been shown in many different models to result in the generation of highly reactive oxygen species (Coyle and Puttfarcken, 1993; Dugan et al., 1995), which are able to attack critical cellular components, including DNA, proteins and phospholipids (Halliwell and Gutteridge, 1985; Chan, 2001).

Several experimental evidences indicate that lipid peroxidation results in loss of membrane integrity, impairment of the function of membrane-transport proteins and ion channels, disruption of cellular ion homeostasis and concomitantly increases neuronal vulnerability to excitotoxicity (Mattson, 1998; Springer et al., 1997). In addition, several studies have demonstrated that glutamate uptake is compromised by pathophysiological events associated with the generation of free radical species (Volterra et al., 1994; Keller et al., 1997; Springer et al., 1997). Dysfunction of the glutamate transporters can lead to neuronal damage by allowing glutamate to remain in the synaptic cleft for a longer duration, contributing to excitotoxicity-induced oxidative injury in the ischemic brain.

Results from different studies indicate that oxidative damage in the gerbil hippocampus is maximal at late stages after ischemia (Hall et al., 1997; Oostveen et al., 1998; Yamaguchi et al., 1998; Urabe et al., 2000; Candelario-Jalil et al., 2001). The delayed occurrence of oxidative stress correlates well with delayed neuronal loss of hippocampal CA1 pyramidal neurons (Hall et al., 1993, 1997; Oostveen et al., 1998), suggesting that reactive oxygen species formation may cooperate in a series of molecular events that link ischemic injury to neuronal cell death.

It is noteworthy that a delayed treatment with the cyclooxygenase-2 inhibitor nimesulide is able to reduce the late-onset oxidative injury in the ischemic hippocampus (Table 1 and Fig. 1). This might suggests that cyclooxygenase-2 is a critical source of free radicals at later times following global cerebral ischemia, since no further protection was observed when this cyclooxygenase-2 inhibitor was administered 30 min before ischemia. Cyclooxygenase-2 is one of a select few proteins that still remains upregulated in CA1 hippocampal cells even at 3 days after ischemia (Nakayama et al., 1998; Koistinaho et al., 1999), thus preceding the death of these neurons.

The production of pro-inflammatory prostanoids is another injurious mechanism associated to the cyclooxygenase-2 enzymatic activity (Marnett et al., 1999). Interestingly, prostaglandins have been shown to stimulate calcium-dependent glutamate release in astrocytes (Bezzi et al., 1998), thus contributing to excitotoxicity. Pharmacological inhibition of COX-2 has been shown to reduce N-methyl-D-aspartate-mediated neuronal cell death both in vitro (Hewett et al., 2000) and in vivo (Iadecola et al., 2001). In addition, recent investigations have found a potentiation of excitotoxicity in transgenic mice overexpressing neuronal COX-2 (Kelley et al., 1999) and a significant reduction in ischemic brain injury in COX-2-deficient mice (Iadecola et al., 2001).

Probably the most important and significant alteration in the antioxidant system is a decrease in reduced glutathione content (Schulz, 2000). Nimesulide treatment was able to confer protection against hippocampal glutathione depletion (Table 1) and the ensuing lipid peroxidation as assessed by the increases in MDA, 4-HDA and lipid hydroperoxides levels. No significant modification in lipid peroxidation was observed in animals treated with nimesulide without ischemia (sham-operated animals; Table 2). This result suggests that administration of cyclooxygenase-2 inhibitors does not modify basal lipid peroxidation in brain possibly due to the low expression of this enzyme under physiological conditions (Seibert et al., 1997). On the contrary, inhibition of cyclooxygenase-2 with nimesulide at 48 h after the ischemic episode, significantly reduced lipid peroxidation and glutathione depletion, indicating that under damaging conditions to the brain, cyclooxygenase-2 contributes to damage through a mechanism involving production of highly reactive free radicals.

The biochemical evidence of reduction of oxidative stress by nimesulide in hippocampus and its neuroprotective efficacy is supported by the histofluorescence findings using the specific fluorescent marker Fluoro-Jade B (Fig. 2). It is important to emphasize that these beneficial effects were seen at doses that are well tolerated in patients with minimal side effects. These findings are in line with other studies, which have shown that cyclooxygenase inhibitors are able to reduce free radical production in cerebral ischemia (Hall et al., 1993; Domoki et al., 2001; Miyamoto et al., 2003) and in traumatic brain injury (Tyurin et al., 2000).

In summary, the results of the present study indicate that nimesulide confers protection against ischemic global brain injury through a mechanism that involves reduction of oxidative damage. Although other mechanisms might also be involved, we believe that inhibition of cyclooxygenase-2 with the concomitant attenuation of oxidative stress and prostaglandin accumulation is the main event accounting for the neuroprotective properties of nimesulide. Our findings are consistent with the hypothesis that reducing oxidative events should prove beneficial in promoting cell function and survival following ischemic brain injury and likely hold a therapeutic promise to intervene neuronal injury evolving after global cerebral ischemia with the cyclooxygenase-2 inhibitor nimesulide.

**Table 1.** Effect of the selective COX-2 inhibitor nimesulide on ischemia-induced glutathione depletion in the hippocampus after 48 h of reperfusion following 5 min global cerebral ischemia.

| Groups | GSH (µg/g tissue) | GSSG (ng/g tissue) |
|---|---|---|
| *Delayed treatment schedule* | | |
| Sham | 1.53 ± 0.19 | 1.52 ± 1.28 |
| Ischemia + Vehicle | 0.99 ± 0.21 $^{**}$ | 10.24 ± 3.16 $^{**}$ |
| Ischemia + Nimesulide (3 mg/kg) | 1.21 ± 0.14 $^{\ddagger *}$ | 6.82 ± 1.94 $^{\ddagger *}$ |
| Ischemia + Nimesulide (6 mg/kg) | 1.35 ± 0.24 $^{\ddagger *}$ | 5.33 ± 2.08 $^{\ddagger *}$ |
| Ischemia + Nimesulide (12 mg/kg) | 1.32 ± 0.18 $^{\ddagger *}$ | 5.71 ± 3.16 $^{\ddagger *}$ |
| *Pretreatment* | | |
| Ischemia + Vehicle | 1.02 ± 0.18 $^{**}$ | 9.84 ± 3.55 $^{**}$ |
| Ischemia + Nimesulide (6 mg/kg) | 1.29 ± 0.26 $^{\dagger *}$ | 5.11 ± 2.17 $^{\dagger *}$ |

Data are mean ± S.D. $^{**}$ $P<0.01$ compared to sham. $^{\ddagger}$ $P<0.05$ compared to vehicle (delayed treatment). $^{*}$$P<0.05$ compared to sham. $^{\dagger}$$p<0.05$ compared to vehicle (pretreatment). ANOVA followed by Student-Newman-Keuls post-hoc test.

**Table 2.** Lack of effect of the selective COX-2 inhibitor nimesulide on oxidative stress parameters in sham-operated animals.

| Groups | GSH (µg/g tissue) | GSSG (ng/g tissue) | MDA + 4-HDA (nmol/mg protein) | Hydroperoxides (nmol/mg protein) |
|---|---|---|---|---|
| Sham + Vehicle | 1.53 ± 0.19 | 1.52 ± 1.28 | 3.1 ± 2.1 | 22.4 ± 2.4 |
| Sham + Nimesulide (3 mg/kg) | 1.48 ± 0.20 | 1.55 ± 0.98 | 2.9 ± 1.9 | 23.3 ± 1.7 |
| Sham + Nimesulide (6 mg/kg) | 1.51 ± 0.17 | 1.49 ± 1.02 | 2.8 ± 1.3 | 22.6 ± 2.1 |
| Sham + Nimesulide (12 mg/kg) | 1.56 ± 0.21 | 1.53 ± 0.97 | 3.0 ± 0.9 | 20.8 ± 1.8 |

Data are mean ± S.D.

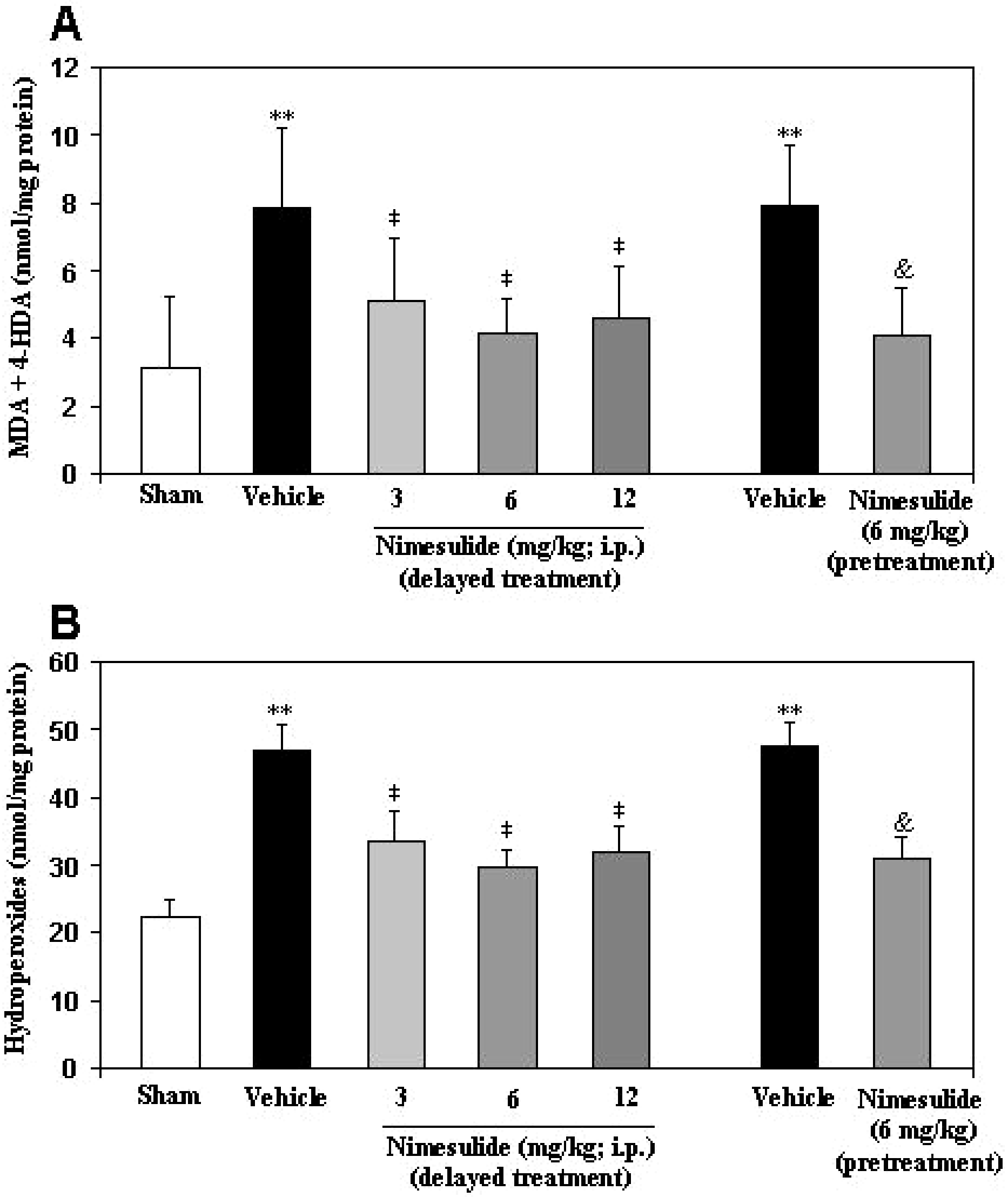

**Fig. 1.** Effect of delayed treatment with different doses of nimesulide on markers of lipid peroxidation in hippocampus at 48 h after 5 min of transient forebrain ischemia in gerbils. Malondialdehyde (MDA) and 4-hydroxy-alkenals (4-HDA) (**A**) and lipid hydroperoxides levels (**B**) were determined as measures of lipid peroxidation. Data are mean ± S.D. **P<0.01 compared to sham. ‡P<0.05 with respect to vehicle-treated animals. &P<0.05 with respect to vehicle (pretreatment). ANOVA followed by Student-Newman-Keuls post-hoc test.

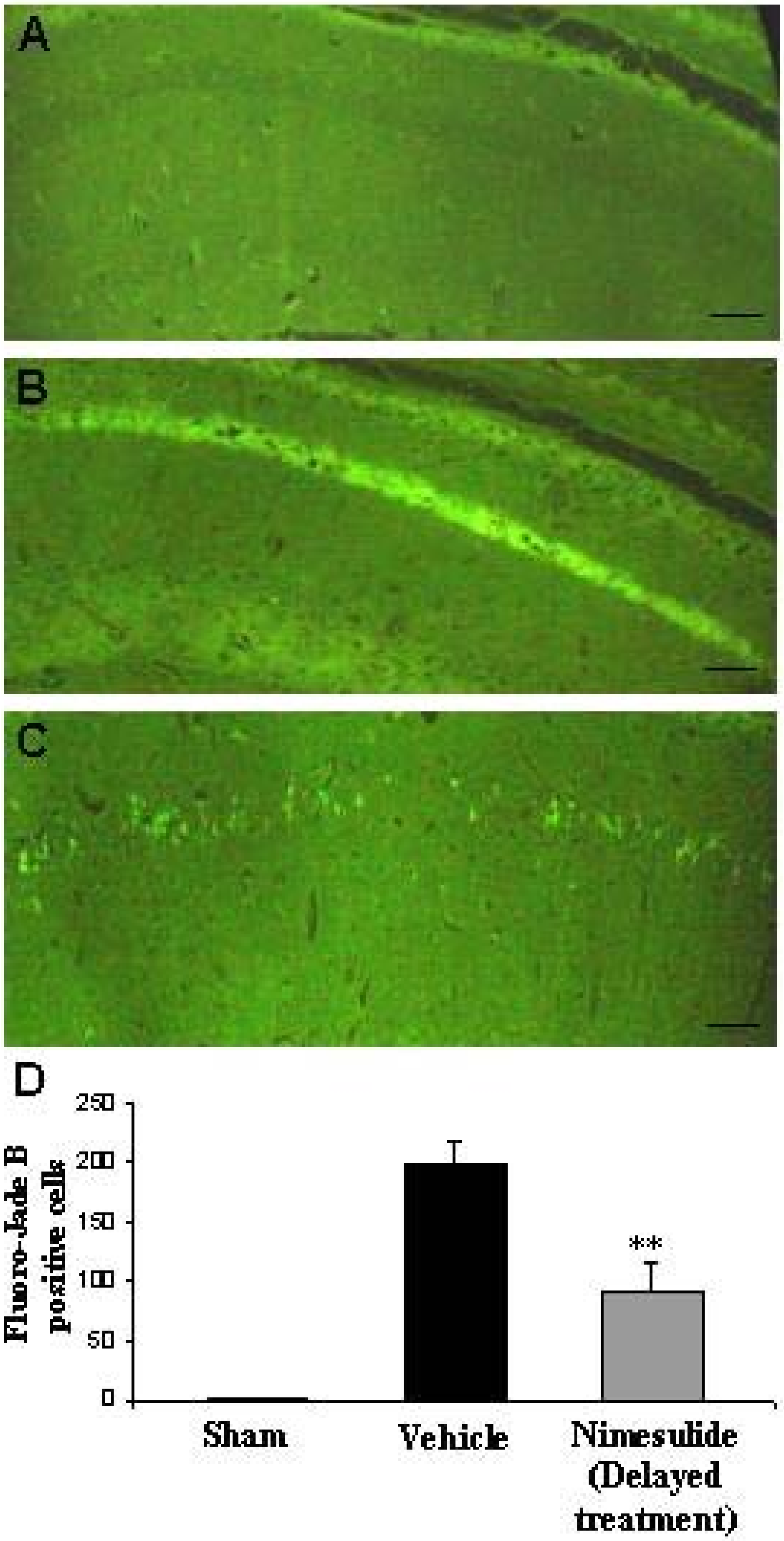

**Fig. 2.** Neuroprotective effect of nimesulide on hippocampal sections as stained with Fluoro-Jade B. This novel reagent labels degenerating cells in yellow, whereas normal cells appear unstained. **A**, CA1 region of hippocampus in the control (sham-operated) animals show no Fluoro-Jade B positive cells. **B**, CA1 region of hippocampus in vehicle-treated ischemic gerbils shows extensive labeling in nearly all pyramidal neurons. **C**, CA1 region of hippocampus in ischemic animals treated with nimesulide (6 mg/kg) shows few Fluoro-Jade B positive cells. **D**, Cell counts of degenerating neurons in the CA1 sector of dorsal hippocampus and neuroprotective effect of the COX-2 inhibitor nimesulide (6 mg/kg; starting after 6 h of reperfusion). Bars represent the group means (+S.D.). **P<0.01 compared with vehicle. Bar =200 μm in panels **A**, **B** and **C**.